\documentclass[pdflatex,sn-nature]{sn-jnl}% Style for submissions to Nature Portfolio journals
\usepackage{graphicx}%
\usepackage{multirow}%
\usepackage{amsmath,amssymb,amsfonts}%
\usepackage{amsthm}%
\usepackage{mathrsfs}%
\usepackage[title]{appendix}%
\usepackage{xcolor}%
\usepackage{textcomp}%
\usepackage{manyfoot}%
\usepackage{booktabs}%
\usepackage{algorithm}%
\usepackage{algorithmicx}%
\usepackage{algpseudocode}%
\usepackage{listings}%
\theoremstyle{thmstyleone}%
\newtheorem{theorem}{Theorem}%  meant for continuous numbers
\newtheorem{proposition}[theorem]{Proposition}% 

\theoremstyle{thmstyletwo}%
\newtheorem{example}{Example}%
\newtheorem{remark}{Remark}%

\theoremstyle{thmstylethree}%
\newtheorem{definition}{Definition}%

\begin{document}

\title[Quantifying Mental States]{Quantifying Mental States in Work Environment: Mathematical Perspectives}


\author[1]{\fnm{Aymen} \sur{Balti}}\email{aymen.balti@univ-lehavre.fr}

\author[1]{\fnm{Assane} \sur{Wade}}\email{assane.wade@univ-lehavre.fr}
\author[1]{\fnm{Abdelatif} \sur{Oujbara}}\email{abdelatif.oujbara@teu.univ-lehavre.fr}
\author[1]{\fnm{M.A.} \sur{Aziz-Alaoui}}\email{aziz.alaoui@univ-lehavre.fr}
\author[2]{\fnm{Hicham} \sur{Bellarabi}}\email{hicham.bellarabi@ca-paris.fr }
\author[2]{\fnm{Fr\'ed\'eric} \sur{Dutertre}}\email{frederic.dutertre@ca-paris.fr  }
\author*[1,3]{\fnm{Benjamin} \sur{Ambrosio}}\email{benjamin.ambrosio@univ-lehavre.fr}

\affil*[1]{ \orgdiv{Department of Mathematics},\orgname{University Le Havre Normandie, Normandie Univ.,
LMAH UR 3821}, \orgaddress{\street{25 rue Philippe Lebon}, \city{Le Havre}, \postcode{76600}, \state{Normandie}, \country{France}}}

\affil[2]{\orgname{Cr\'edit Agricole Ile de France}, \orgaddress{\street{26 quai de la rap\'ee}, \city{Paris}, \postcode{75012}, \country{France}}}

\affil*[3]{ \orgname{The Hudson School of Mathematics}, \orgaddress{\street{244 Fifth Avenue}, \city{New York}, \postcode{10001}, \state{New York}, \country{USA}}}


\abstract{We introduce a novel framework for quantifying mental and emotional states over time by combining virtual reality (VR) exposure with EEG recordings. Participants experienced a stress-inducing work scenario in VR, originally designed as a training tool for bank employees, providing a controlled proxy for high-stakes situations. This setup enables integration of subjective emotional self-assessments with objective neural data, from which an algorithm was efficiently used to infer emotional states. Building on these measurements, we propose possible mathematical models to capture the temporal dynamics of mental states, offering a quantitative approach to studying emotional processing and informing adaptive training in complex environments.
}




\maketitle

\section{Introduction}\label{sec1}
Emotional states continuously emerge during our daily activities. Emotions and their implication in behavior have been a topic of research forever and have been analyzed since ancient civilizations through many prisms over time. This goes from philosophy to politics, from moral and religion to law \cite{Eus2012}. Modern paradigms have initially been shaped by psychology and psychiatry. With advances in technology and aggregation of large data sets, biochemistry and neuroscience, supported by data analysis, provide now a data-driven scientific framework \cite{Ado2018,Bar2015,Dam2013,Ekm2007,Freud1990,Led2000}. In a previous contribution,  wishing to propose a mathematical framework for a descriptive approach of how feelings arise in many aspects of human daily lives, we represented the human mental state as a dynamical system: the mental state is akin to a physical state in which emotional fluxes arise. This mental state evolves as a function of time and results from external stimulus and internal reactions, see \cite{Amb2021}. In several aspects, our model encompasses and goes beyond certain psychological theories such as the Discrete Theory of Emotions \cite{Ekm1971,Ekm2006,Ekm2007}, the Dimensional theory \cite{Cac1986,Mat2017,Pos2005,Rus1980,Sch1952} and the Affective Events Theory \cite{Ash2005,Tru2015,Wei1996}. The affective events theory particularly focuses on the affects triggered by events at work. A fundamental question is how to measure these affects. In the present article, we communicate about the results of a study that involved 87 participants who went through a work-type event on a virtual reality (VR) headset. These subjects were asked to visualize an interactive scenario in which they played the role of a bank employee confronted by an angry client. From time to time, they had to select between few possible responses to the client's behavior, which then determined how the scenario would unfold.  At the end, they were given a grade within the VR headset. While participants engaged in the experience using the headset, we recorded EEG signals using 4 channels (openBCI electrodes, \cite{OpenBCI,Jia2019,Mon2023}).  We also recorded electrodermal activity, Photoplethysmography (known as PPG which is used to derive heart rate) and temperature  thanks to the emotibit framework \cite{Mon2023,Emotibit}. Finally, subjects were asked to answer a questionnaire at the end of the experiment. The scenario was initially developed at the Cr\'edit Agricole Paris for internal training purpose. The aim of the study is to establish a realistic framework for recording data on psychological states, to support the development of dynamical systems models that describe their evolution over time.

\begin{center}

\begin{figure}

\centering
\includegraphics[scale=0.3]{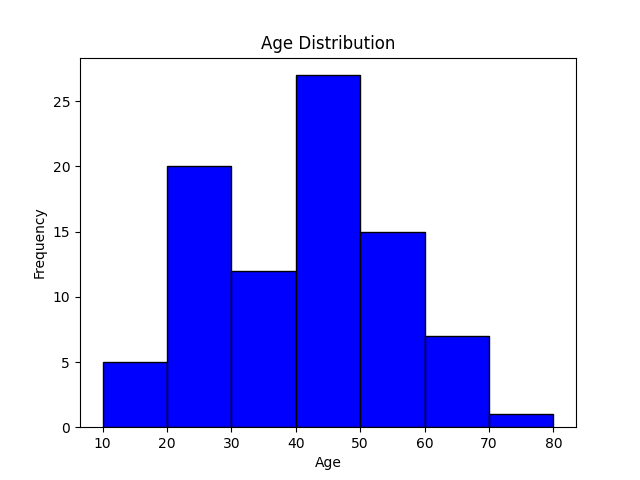}
\includegraphics[scale=0.3]{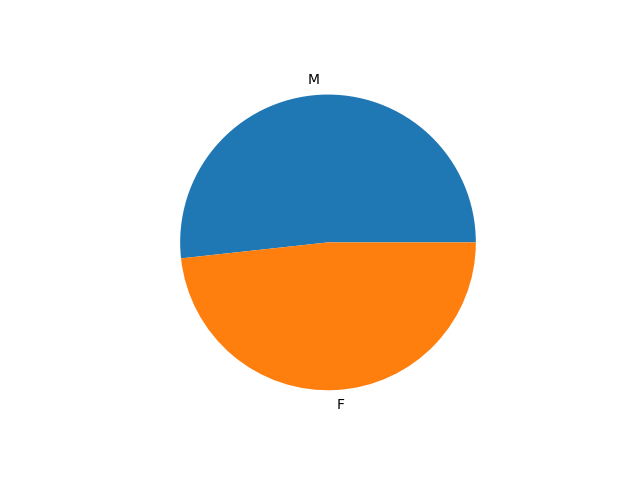}
\caption{Age and Sex Distributions of the sample of participants. We observe a well distributed cohort in age with a pic in ranges 40-50 and 20-30. There was 45 males and 42 females.}
\label{fig:AgeDistribution}
\end{figure}
\end{center}

Participants were recruited in France, in the regions of Paris-Ile de France and Normandie. The age (mean=40.6, std=13.7) and sex (M=45, F=42) distributions are represented in Figure \ref{fig:AgeDistribution}. 

The analysis of the experiments includes the following different parts: 
\begin{itemize}
     \item[--] a qualitative and statistical examination of the participants' responses;
     \item[--] an automatic analysis of the questionnaire responses that generated a valence score: the questionnaire included several items designed to evaluate participants’ internal states during the experiment, including an open-ended question inviting them to freely describe the thoughts, reasoning processes, and emotions they experienced. Based on these responses, we developed an algorithm that assigns each participant a score reflecting the overall positivity or negativity of their reported feelings. This score is represented as a binary variable in ${0,1}$, which can be interpreted as a discrete operationalization of the concept of valence commonly discussed in the psychological literature, see \cite{Cac1986,Mat2017,Pos2005,Rus1980,Sch1952};
     \item[--]  an analysis of the EEG signals: EEG data were preprocessed and analyzed using localized spectral frequencies techniques;
     \item[--] a machine learning algorithm aimed at predicting the valence number from the EEG;
     \item[--] a reflection on mathematical models to be used to produce relevant signals.
\end{itemize}

\section{Results}\label{sec2}
\subsection{Analysis of the questionnaire and Generation of the valence number}

\begin{figure}

\centering
\includegraphics[scale=0.3]{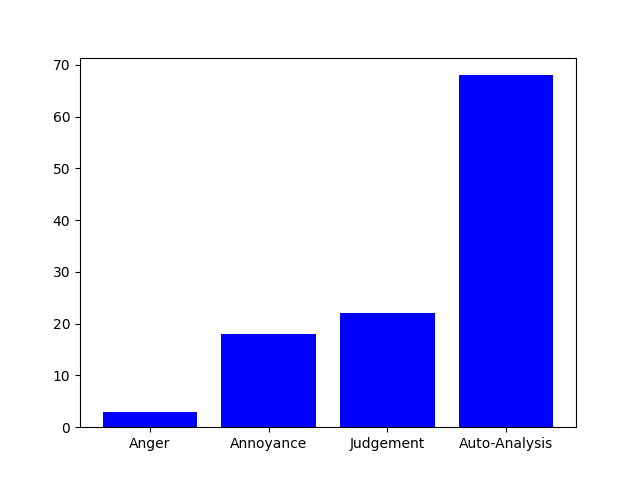}

\caption{This figure summarizes some of the answers to the questionnaire following the VR experiment: 3 of 87 participants (3.4 \%) reported experiencing anger during the experiment, while 18 participants (20.7\%) reported feelings of annoyance and 22 participants (25.3\%) reported a sense of judgment toward the client; 68 declared to be used to analyze their emotions and internal psychological states. These numbers are relatively low, highlighting that most participants experienced only mild emotions in the VR scenario—far less intense than they would likely feel in real life. This aligns with the purpose of the training, which is to allow reflection on challenging situations in a controlled environment, helping participants prepare for real-life interactions without being overwhelmed by strong emotions. }
\label{fig:AnswerToQuestionnaire}
\end{figure}

\begin{figure}
\centering
\includegraphics[scale=0.3]{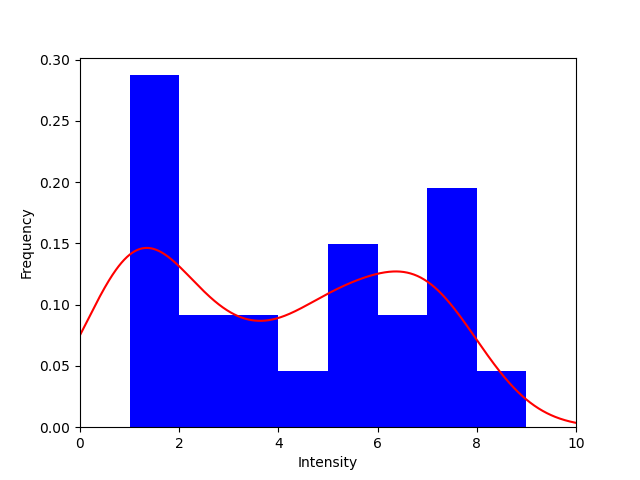}

\caption{This figure illustrates the statistical distribution of responses to the question: “On a scale from 0 to 10, to what extent do you think interacting with the client would have affected your internal state if it had occurred in real life?” The bimodal distribution indicates heterogeneity in participants’ anticipated emotional reactivity, with some expecting minimal impact and others anticipating a more substantial effect. Notably, the anticipated emotional reactivity is much higher than the 3.4\% of anger actually reported in the VR setting. This finding highlights the benefit of VR as a safer environment for exploring challenging situations and preparing for real-life interactions.  }
\label{fig:AnswerToQuestionnaire2}
\end{figure}

Analysis of the questionnaire responses revealed that 3 of 87 participants (3.4 \%) reported experiencing anger during the experiment, while 18 participants (20.7\%) reported feelings of annoyance and 22 participants (25.3\%) reported a sense of judgment toward the client (see figure \ref{fig:AnswerToQuestionnaire}). We note that although the client’s behavior in VR is deliberately designed to provoke emotions such as anger— if it was real life - only three participants reported feeling anger during the VR experience. These numbers suggest that, although the scenario was designed to provoke strong negative emotions if it was in real life, the majority of participants experienced only mild to moderate affective reactions, which indicates a degree of emotional distancing afforded by the VR setting. Indeed, participants are aware they are not in a real-life situation while immersed in VR during this experiment. This distinction makes the VR experiment particularly suitable for training and educational purposes, for which it was designed: individuals can learn to recognize and reflect on potential scenarios without being overwhelmed by the critical emotional states they might experience in reality. In this way, VR allows participants to reason through situations in a controlled environment, better preparing them for real-world encounters where emotions are more likely to arise. When asked  ``On a scale from 0 to 10, to what extent do you think the interaction with the client would have affected your internal state if it had occurred in real life?", participants reported a mean score of 4.0 (SD = 2.5). This distribution, illustrated in Figure \ref{fig:AnswerToQuestionnaire2}, indicates a moderate perceived impact on internal state, with considerable variability across participants. It was not uniform but displayed a bimodal tendency, with clusters around low values (1–2) and higher values (6–8). This pattern suggests heterogeneity in participants’ anticipated emotional reactivity, with some perceiving minimal potential impact and others anticipating a more substantial effect. However, it appears to be much more higher that the declared 3.4\% of anger in the VR setting. The questionnaire further examined whether participants engaged in regular self-monitoring of their emotional states, a factor that may systematically influence response patterns; 68 participants (78 \%) affirmed that they habitually reflect on their emotions. Finally, participants were asked to freely describe the thoughts, reasoning processes, and emotions that arose during the experiment. A selection of responses is presented below: \textit{``I kept my temper in front of the customer''; ``The woman spoke badly to me''; ``I felt empathy. I was worried about the customer’s case. At the same time, I was annoyed by the customer’s behavior because she was not kind, even though I suggested a solution''; ``I have been in this situation many times. I am able to control my emotions in such contexts because I have worked in this field for more than 20 years"; ``I felt stress at the beginning, but as soon as I understood her needs, I was able to manage the situation"}.

Our algorithm, when applied to the questionnaire responses, assigned each participant a valence score—0 indicating predominantly negative feelings and 1 indicating predominantly positive feelings regarding the experiment. Of the 87 participants, 60 (69.0\%) were classified with a positive valence (1), while the remaining 27 (31.0\%) were classified with a negative valence (0). See figure \ref{fig:AnswerToQuestionnaire2}. These findings indicate that participants tend to have a positive experience during the VR scenario, despite the fact that the event would be emotionally demanding if encountered in real life.

\begin{figure}

\centering
\includegraphics[scale=0.3]{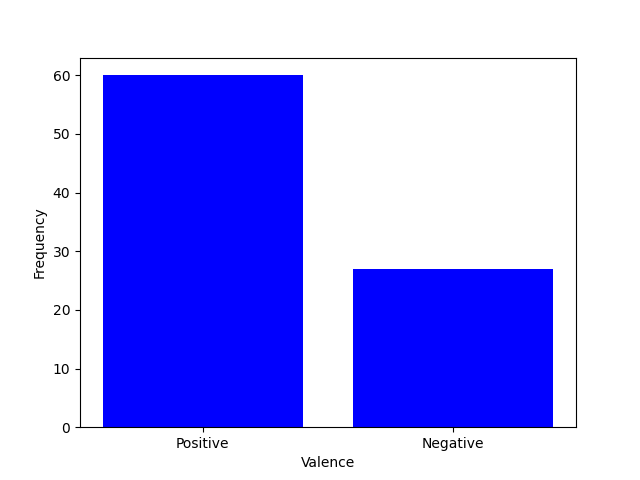}

\caption{Frequencies of positive and negative valences assigned by our algorithm to each participant's feelings from the responses to the questionnaire.  Of the 87 participants, 60 (69.0\%) were classified with a positive valence (1), while the remaining 27 (31.0\%) were classified with a negative valence (0). These findings indicate that participants tend to have a positive experience during the VR scenario, despite the fact that the event would be emotionally demanding if encountered in real life.  }
\label{fig:ValenceFromQuestionnaire}
\end{figure}

\subsection{EEG Analysis and Prediction of Valence from EEG}
\begin{figure}

\centering
\includegraphics[scale=0.2]{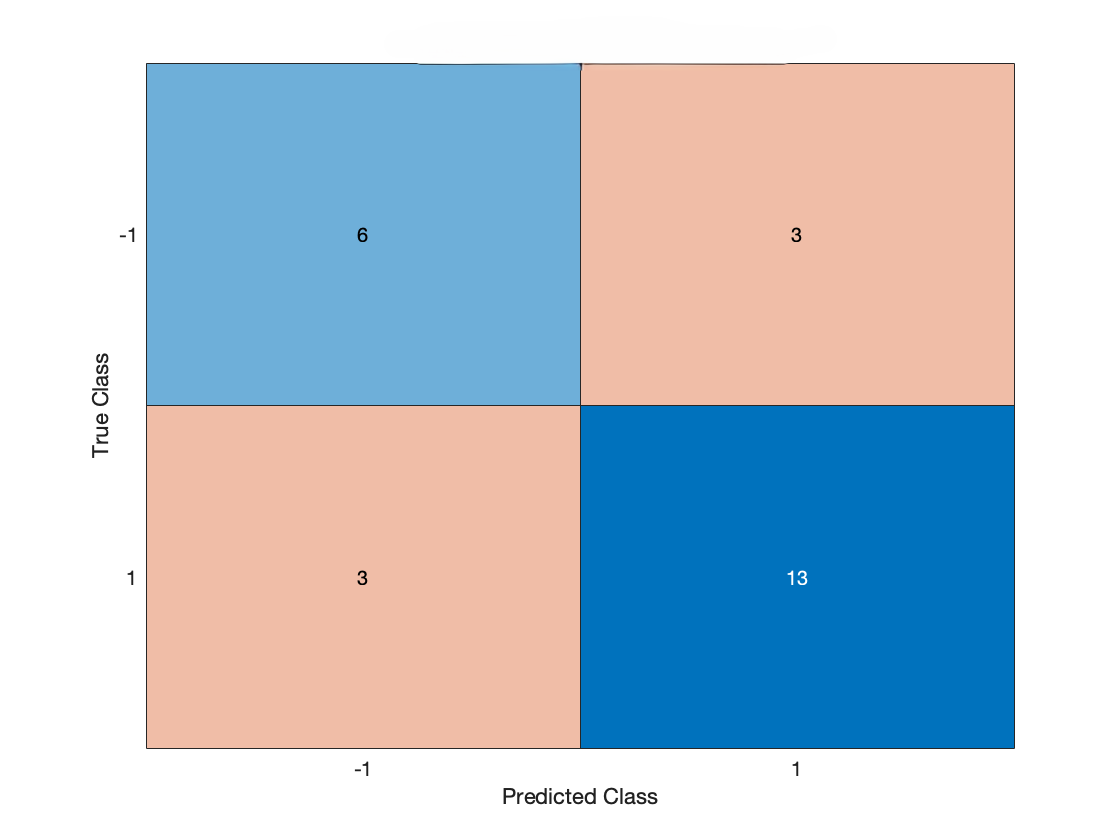}

\caption{The KNN algorithm predicted valence in the test set with an accuracy of 81.2\% for valence $= 1$ and 66.7\% for valence $= 0$.}
\label{fig:Prediction}
\end{figure}

Next, we analyzed the EEG data from each participant and extracted spectral features within sub-time windows. These features were then used to train a K-Nearest Neighbors (KNN) algorithm to predict valence using 59 of the participants. The KNN algorithm predicted valence in the test set (25 participants) with an accuracy of 81.2\% for valence $= 1$ and 66.7\% for valence $= 0$. Figure \ref{fig:Prediction} summarizes the results.

\subsection{Generating relevant signals with Mathematical Models}
\begin{figure}

\centering
\includegraphics[scale=0.3]{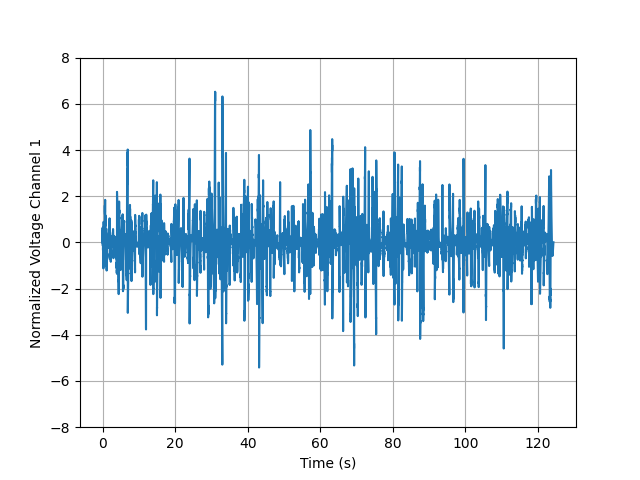}
\includegraphics[scale=0.3]{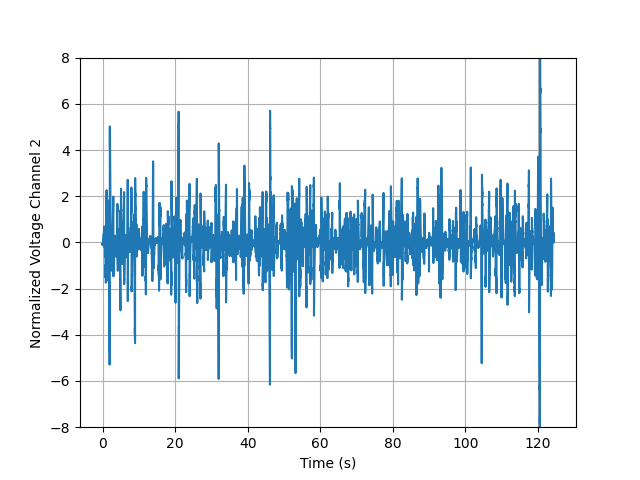}\\
\includegraphics[scale=0.3]{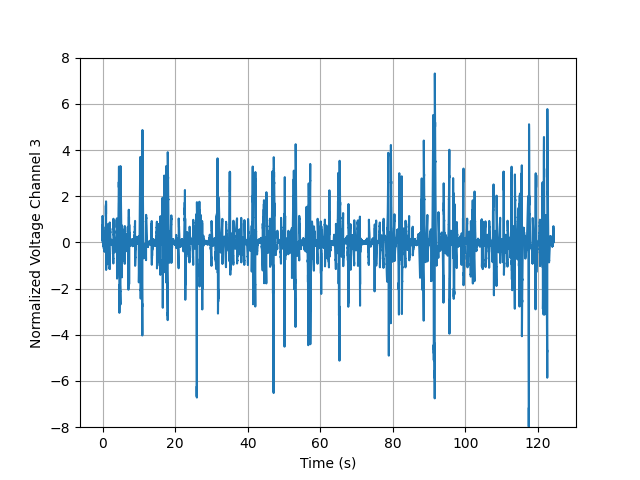}
\includegraphics[scale=0.3]{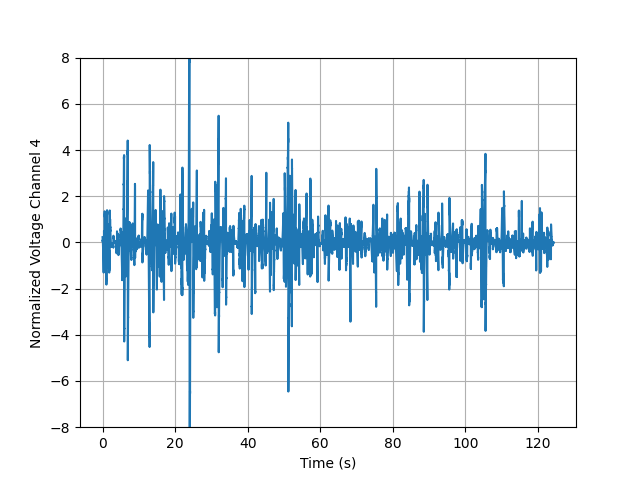}\\
\includegraphics[scale=0.3]{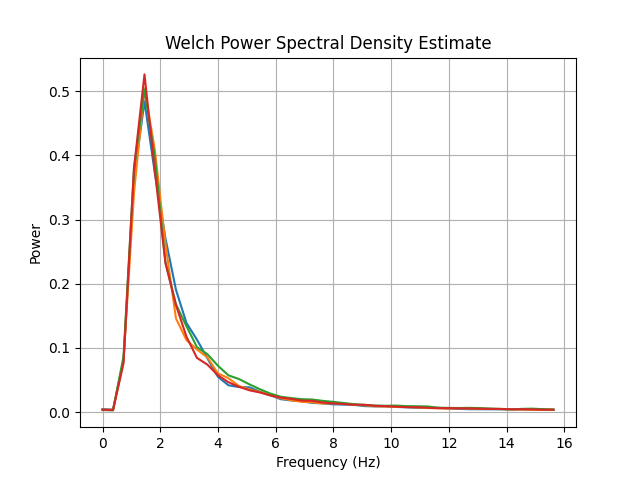}
\includegraphics[scale=0.3]{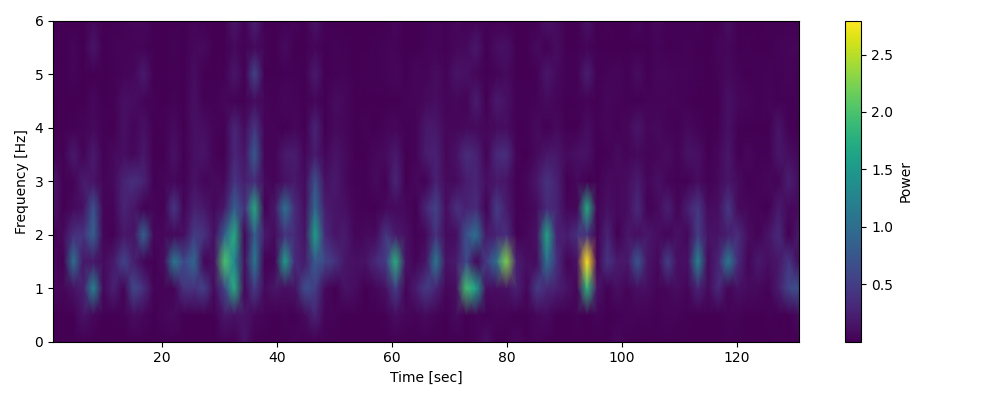}
\caption{EEG data recorded from one of the participants. The top two rows display the preprocessed signals from four recording channels. The lower-left panel shows the corresponding power spectral densities (PSDs), and the lower-right panel presents the spectrogram of the first channel. We remark that the preprocessed signals of the four channels present a pic at around 1.5 Hz with a maximal power of 0.5.}
\label{fig:EEG-Data}
\end{figure}

Figure \ref{fig:EEG-Data} displays the preprocessed EEG time series of one participant alongside a spectral analysis of the same data. The extracted spectral features served as inputs to the prediction model. As a natural next step, we address the question of the ability of mathematical models to reproduce relevant signals using mathematical models. For this purpose, several approaches, including neural networks driven by Hodgkin–Huxley equations, a reduced stochastic model, reaction–diffusion models, and a forced Poisson equation can be considered. Figure \ref{fig:EEG-Data} shows the simulation results of the reduced stochastic model, which is a stochastic differential equation derivated from a  FitzHugh-Nagumo model type. The relevance of this model lies in its ability to produce a recurrent switch between two attracting states, as observed in the raw data (not shown here), which could relate to the EEG microstate paradigm \cite{Mic2018}. A more comprehensive study—covering parameter estimation and comparison with experimental data—will be presented in a forthcoming article. Additional details are provided in the Results and Discussion section.

\begin{figure}
\label{fig:FHNSto}
\centering
\includegraphics[scale=0.3]{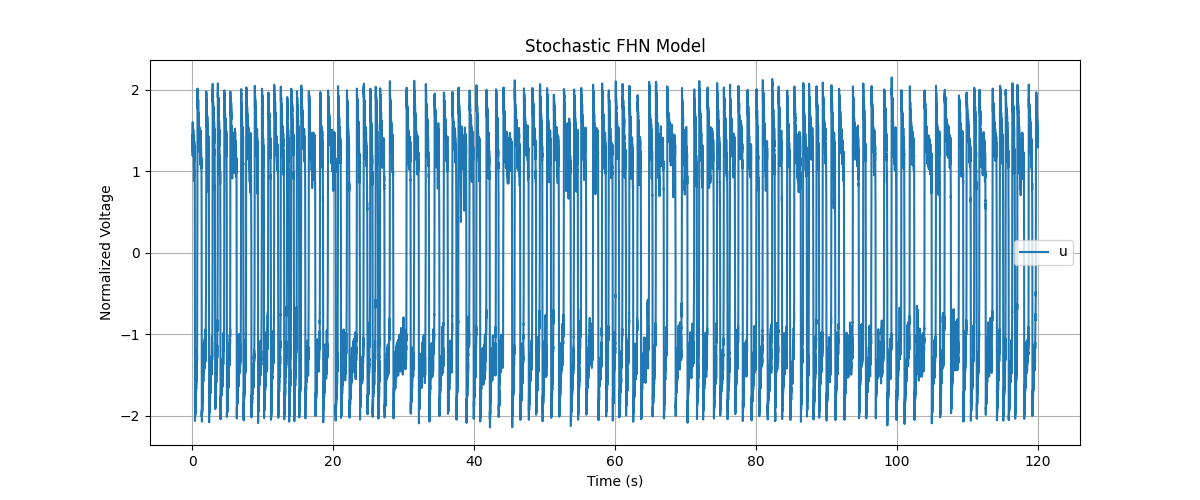}
\includegraphics[scale=0.3]{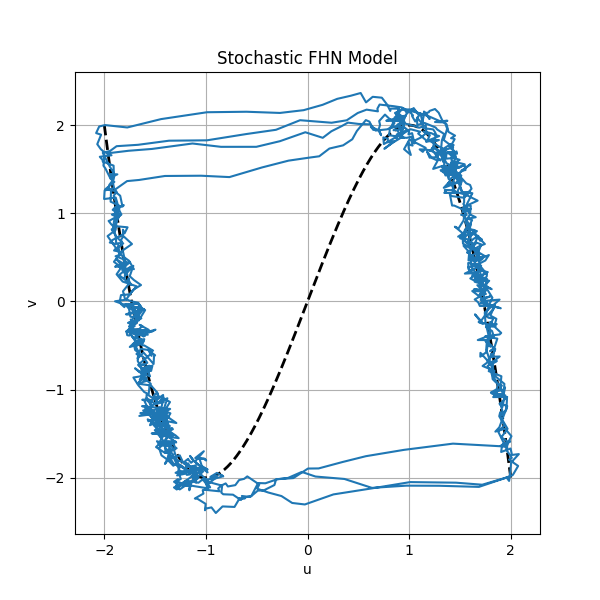}
\includegraphics[scale=0.3]{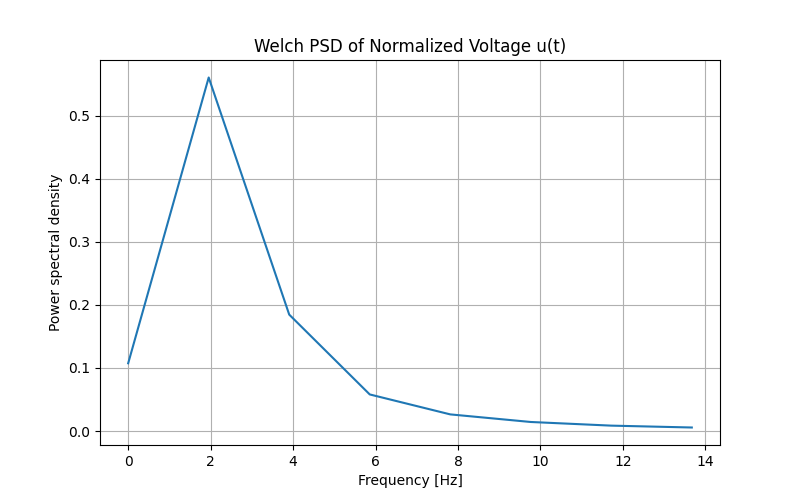}

\caption{Numerical simulation of a SDE of FHN type. The top panel shows the time series of the fast variable $u$. The lower-left panel depicts the corresponding phase-space trajectory over a five-second interval, while the lower-right panel presents the power spectral density (PSD) of the rescaled signal $
1.6u$. We recover the pic at around 1.9 with a maximal power around 0.5 which is close to the characteristics of the preprocessed signals displayed in figure \ref{fig:EEG-Data}}
\end{figure}

\section{Methods}
Experiments were conducted using a Vive HTC virtual reality headset. Questionnaire responses were processed using a custom MATLAB implementation based on thresholding. For each response series, a scalar score was derived and subsequently classified as positive or negative according to a predefined cutoff criterion, yielding an operational measure of emotional state during the experiment.
Electroencephalographic (EEG) signals were recorded with the OpenBCI Ganglion Board (4-channel, 200 Hz sampling rate), following the international 10–20 electrode placement system.  Active electrodes were positioned at Fp1 and Fp2 (frontopolar sites) and T3 and T4 (temporal sites). Reference and ground electrodes were placed at A1 and A2 (left and right earlobes, respectively). The raw signals were digitized onboard and transmitted wirelessly to the acquisition computer. Preprocessing included band-pass filtering between 1–40 Hz (Butterworth, 4th order), notch filtering at 50 Hz to suppress line noise, and visual inspection to reject epochs contaminated by motion or muscle artifacts. Data were normalized prior to analysis. For spectral analysis, power spectral densities (PSDs) were estimated using Welch’s method with 2-second Hamming windows, 75\% overlap. For the classification of EEG signals, we employed the k-nearest neighbors (KNN) algorithm using the spectral features extracted from four EEG channels. The model was trained on data from 62 participants and subsequently evaluated on an independent set of 25 participants. Several mathematical models can be employed to reproduce signals with relevant spectral properties in EEG context. In this context, our own repertoire includes networks of excitatory and inhibitory Hodgkin-Huxley (HH)–type neurons \cite{Maa2024}, reaction-diffusion systems \cite{Amb2009,Amb2012,Amb2013,Amb2016,Amb2017,Amb2017-2,Amb2023,Amb2024}, time-driven Poisson equations such as in \cite{Xi2022} and reduced stochastic dynamical systems \cite{Amb2022}. We provide below representative examples of each class of these systems. The following equations (\cite{Maa2024}) describe the time evolution of a network of 
$N$ Hodgkin–Huxley (HH) neurons, with additional terms accounting for excitatory and inhibitory presynaptic inputs:

\begin{equation}\label{eq:HH-Network}
\left\{
\begin{array}{rcl}
C\dfrac{d}{dt} V_i &=& \overline{g}_{Na} m_i^3 h_i(E_{Na}-V_i) + \overline{g}_{K} n_i^4(E{K}-V_i) + \overline{g}_{L}(E_{L}-V_i) \\
& & + g_{Ei}(E_{E}-V_i) + g_{Ii}(E_{I}-V_i), \quad i \in {1,\dots,N} \\
\dfrac{d}{dt} x_i &=& \alpha_x(V_i)(1-x_i) - \beta_x(V_i)x_i, \quad x \in {m,n,h} \\
\tau_E \dfrac{d}{dt} g_{Ei} &=& -g_{Ei} + S^{dr}\sum_{s\in \mathcal{D}(i)}\delta(t-s) + S^{QE}\sum_{j\in \Gamma_E(i),, s \in \mathcal{N}(j)}\delta(t-s) \\
\tau_I \dfrac{d}{dt} g_{Ii} &=& -g_{Ii} + S^{QI}\sum_{j\in \Gamma_I(i), s \in \mathcal{N}(j)}\delta(t-s)
\end{array}
\right.
\end{equation}
Here, $Q=E$ for excitatory neurons and  $Q=I$ for inhibitory neurons.The first four equations correspond to the canonical HH formalism (membrane voltage and gating variables) augmented with additional terms representing excitatory and inhibitory fluxes from presynaptic neurons and a stochastic external drive. The last two equations describe the synaptic conductances driven by stochastic input and recurrent connectivity. The framework admits substitution of the HH formalism by alternative single-neuron models, and additional terms can be introduced to capture time-dependent external events. Convolution of the network activity (\cite{Maa2024}) provides a global signal reflecting emergent population-level dynamics that can be calibrated against observed EEG recordings.
Another class of models are Reaction-Diffusion systems which write:
\begin{equation}
\label{eq:RD}
U_t=F(U)+D\Delta U+c(x,t)
\end{equation}

Here, $U(x,t)$ denotes the state variables of interest, such as membrane voltage or gating variables, while $\Delta U$ captures spatial diffusion, and $c(x,t)$ represents source terms relevant for reproducing EEG activity. These sources may include both deterministic and stochastic inputs; see, for example, \cite{Vol2024}. Again, the resulting aggregate spatial activity can then be calibrated against empirical EEG recordings.

The standard modeling strategy is to introduce a non-autonomous Poisson equation, derived from Maxwell’s equations:

\begin{equation}
\left\{
\begin{array}{rl}
\nabla \cdot \mathbf{E} &= \frac{\rho}{\varepsilon_0} \\[6pt]
\nabla \cdot \mathbf{B} &= 0 \\[6pt]
\nabla \times \mathbf{E} &= -\frac{\partial \mathbf{B}}{\partial t}  \\[6pt]
\nabla \times \mathbf{B} &= \mu_0 \mathbf{J} + \mu_0 \varepsilon_0 \frac{\partial \mathbf{E}}{\partial t} 
\end{array}
\right.
\end{equation}

where $\rho$ is the free charge density and $\mathbf{J}_{\text{tot}}$ is the total current density.
Applying the quasi-static approximation and further steady state assumptions yields the general form
\begin{equation}
  -\nabla \cdot \big( D(\mathbf{x},t) \nabla u(\mathbf{x},t) \big)
  = F(\mathbf{x},t) + \eta(\mathbf{x},t),
\end{equation}
where $u(\mathbf{x},t)$ is the cortical potential field associated with EEG measurements, $D(\mathbf{x},t)$ is a time-dependent conductivity or diffusion coefficient, $F(\mathbf{x},t)$ encodes non-autonomous source terms (e.g., stimuli, tasks, or inputs), and $\eta(\mathbf{x},t)$ accounts for stochastic fluctuations. In its simplest form, the model reduces to
\begin{equation} 
  -\Delta U = F(\mathbf{x},t),
\end{equation}
providing a tractable approximation; we refer also the reader to \cite{Fra2025} for a recent critical analysis of the derivation. $F(x,t)$ can represent neuronal activity and can be generated by another neuronal model such as \eqref{eq:HH-Network} or \eqref{eq:RD}, with simplified reaction terms. See \cite{Amb2009,Amb2016,Amb2022} for examples in which neuronal activity is mimicked thanks to a FitzHugh-Nagumo \cite{Fit1961,Nag1962} type equation.

The signals presented in the Results section, \ref{fig:FHNSto} were obtained from simulations of the following stochastic differential equation (SDE):

\begin{equation}
\label{eq:FHNSto}
\left\{\begin{array}{rcl}
\epsilon du_t&=&f(u_t)-v_t+\epsilon\sigma_1dB^1_t\\ 
dv_t&=& bu_t-v_t+\sigma_2dB^2_t
\end{array}
\right.
\end{equation}
with $f(u)=-u^3+a^2u$, $a=\sqrt{3}$, $b=1.8$, $\sigma_1=\sigma_2=0.5$.

For these parameters, the deterministic counterpart of \eqref{eq:FHNSto} admits two stable stationary points, $(\sqrt{a^2-b},b\sqrt{a^2-b})$ 
 together with an unstable equilibrium at $(0,0)$. The two stable points lie near the local maxima of the cubic function $-u^3+a^2u$
 along its stable manifold. The stochastic perturbations introduced by the Brownian terms induce random transitions between the two stable branches of the cubic nullcline, resulting in intermittent switching behavior. The average time between two successive switches is approximately $2.77s$, giving rise to the characteristic frequency observed in the simulations.
 Proposition~\ref{prop:FHNbistable} and  slow–fast analysis techniques\cite{Fen1979,Jon1995,Kue2015,Rin1985,Amb2021-2} provide the theoretical background underlying the phenomenon. 

 For small noise intensities, Freidlin–Wentzell large deviation theory provides analytical and computational tools \cite{fre2012,ber2004,ber2006,wei2004,hey2008,hey2015,kra1940,han1990,nei1987,rit2016} to estimate escape times from the stable equilibria. Although in the present application the noise amplitude cannot be considered small, the qualitative mechanism of noise-induced switching remains highly relevant. Such state-switching dynamics are particularly meaningful for interpreting microstate transitions observed in EEG recordings \cite{leh1987,hay2025,kas2025}. Beyond EEG analysis, this modeling framework can also be viewed as a conceptual representation of mental state alternations—between more automatic, emotion-driven modes and more controlled, introspective states. Figure~\ref{fig:FHNSto} further demonstrates that the model can be calibrated to reproduce frequencies observed in empirical data.

\begin{proposition}
\label{prop:FHNbistable}
The underlying deterministic system of \eqref{eq:FHNSto} admits three stationary solutions:
\[U^*_1=(0,0), U^*_2=(u^*,v^*), U^*_3=-U^*_2,\]
where $(u^*,v^*)$ are solutions of
\[f(u^*)-bu^*=0, v^*=bu^*.\]
Assume $\epsilon>0$ is small enough, then, $U^*_1$ is a saddle node whereas $U^*_2$ and $U^*_3$ are stable nodes.
\end{proposition}

Here, we present only the simulation of this model; future work will focus on calibrating the models against EEG and including Emotibit data.
\section{Discussion}\label{sec12}
In this study, we analyzed EEG data from a cohort of 87 participants exposed to a stressful work-related scenario presented through a virtual reality headset. Emotional valence during the task was assessed using both self-report questionnaires and a custom threshold-based algorithm that classified responses as positive or negative. Spectral features were extracted from the EEG recordings and subsequently used to train a k-nearest neighbors (KNN) classifier. The model was trained on data from 59 participants and evaluated on an independent test set of 25 participants, achieving an accuracy of 81.2\% for positive valence (valence = 1) and 66.7\% for negative valence (valence = 0). Beyond empirical analysis, we also explored mathematical models capable of reproducing signals with spectral characteristics comparable to those observed in the experimental data, thereby providing a framework for linking theoretical modeling with empirical findings. Two important remarks are necessary when interpreting these results. First, the VR-based scenarios, although effective in creating controlled experimental conditions, cannot fully replicate the affective dynamics of real-life situations. For example, the proportion of participants reporting anger was only 3.4\%, which is considerably lower than would be expected in naturalistic contexts or even when viewing emotionally evocative film material \cite{Das2023}.. This discrepancy likely reflects the original purpose of the VR scenarios, which were designed for employee training rather than affect induction, leading participants to adopt a more analytical than experiential stance during the task. Second, the EEG recordings were limited to four electrodes, which restricts the spatial resolution of the neural data. Despite these constraints, the classification algorithm performed robustly, achieving a good accuracy in distinguishing emotional valence, underscoring both the feasibility and the potential of combining EEG-based spectral features with machine learning for dynamic emotion recognition, see \cite{Liu2023}. Then we discussed simplified mathematical models capable of reproducing signals indicative of neural electrical activity and corresponding emotional states. Future work will involve a systematic comparison between these simulated signals and empirical EEG recordings, as well as the integration and analysis of data acquired from EmotiBit devices.
% \cite{Law2025} \cite{Men2025} COMBINE FORWARD PROBLEM AND INVERSE PROBLEM.

\textbf{Funding}\\
This study was part of the emergent projet MME funded by the Normandie R\'egion. We thank also the Cr\'edit Agricole Ile de France and the Hudson School of Mathematics for material support. 

\begin{thebibliography}{10}
\expandafter\ifx\csname url\endcsname\relax
  \def\url#1{\burl{#1}}\fi
\expandafter\ifx\csname urlprefix\endcsname\relax\def\urlprefix{URL }\fi
\providecommand{\bibinfo}[2]{#2}
\providecommand{\eprint}[2][]{\url{#2}}
\providecommand{\doi}[1]{\url{https://doi.org/#1}}
\bibcommenthead

\bibitem{Eus2012}
\bibinfo{author}{Nicole, E.} \emph{et~al.}
\newblock \bibinfo{title}{Ahr conversation: The historical study of emotions}.
\newblock \emph{\bibinfo{journal}{The American Historical Review}} \textbf{\bibinfo{volume}{117}}, \bibinfo{pages}{1487–1531} (\bibinfo{year}{2012}).

\bibitem{Ado2018}
\bibinfo{author}{Adolphs, R.} \& \bibinfo{author}{Anderson, D.}
\newblock \emph{\bibinfo{title}{The Neuroscience of Emotion. A new synthesis}}  (\bibinfo{publisher}{Princeton University Press}, \bibinfo{year}{2018}).

\bibitem{Bar2015}
\bibinfo{author}{Barrett, L.} \& \bibinfo{author}{Russell, J.}
\newblock \emph{\bibinfo{title}{The Psychological Construction of Emotion.}}  (\bibinfo{publisher}{New York: NY: Guildofrd Press.}, \bibinfo{year}{2015}).

\bibitem{Dam2013}
\bibinfo{author}{Damasio, A.} \& \bibinfo{author}{Carvalho, G.~B.}
\newblock \bibinfo{title}{The nature of feelings: evolutionary and neurobiological origins}.
\newblock \emph{\bibinfo{journal}{Nature Reviews Neuroscience}} \textbf{\bibinfo{volume}{14}}, \bibinfo{pages}{143--152} (\bibinfo{year}{2013}).

\bibitem{Ekm2007}
\bibinfo{author}{Ekman, P.}
\newblock \emph{\bibinfo{title}{Emotions Revealed}}  (\bibinfo{publisher}{Howl Books}, \bibinfo{year}{2007}).

\bibitem{Freud1990}
\bibinfo{author}{Freud, S.}
\newblock \emph{\bibinfo{title}{The Ego and The Id}}  (\bibinfo{publisher}{W. W. Norton \& Company, New York}, \bibinfo{year}{1989}).

\bibitem{Led2000}
\bibinfo{author}{LeDoux, J.~E.}
\newblock \bibinfo{title}{Emotion circuits in the brain}.
\newblock \emph{\bibinfo{journal}{Annual Review of Neuroscience}} \textbf{\bibinfo{volume}{23}}, \bibinfo{pages}{155–184} (\bibinfo{year}{2000}).

\bibitem{Amb2021}
\bibinfo{author}{Ambrosio, B.}
\newblock \bibinfo{title}{Beyond the brain: towards a mathematical modeling of emotions}.
\newblock \emph{\bibinfo{journal}{Journal of Physics: Conference Series}} \textbf{\bibinfo{volume}{2090}}, \bibinfo{pages}{012119} (\bibinfo{year}{2021}).

\bibitem{Ekm1971}
\bibinfo{author}{Ekman, P.} \& \bibinfo{author}{Friesen, W.~V.}
\newblock \bibinfo{title}{Constants across cultures in the face and emotion.}
\newblock \emph{\bibinfo{journal}{Journal of Personality and Social Psychology}} \textbf{\bibinfo{volume}{17}}, \bibinfo{pages}{124--129} (\bibinfo{year}{1971}).
\newblock \urlprefix\url{https://doi.org/10.1037/h0030377}.

\bibitem{Ekm2006}
\bibinfo{author}{Ekman, P.}
\newblock \bibinfo{title}{Emotions inside out-introduction}.
\newblock \emph{\bibinfo{journal}{Annals of the New York Academy of Sciences}} \textbf{\bibinfo{volume}{1000}}, \bibinfo{pages}{1--6} (\bibinfo{year}{2006}).
\newblock \urlprefix\url{https://doi.org/10.1196/annals.1280.002}.

\bibitem{Cac1986}
\bibinfo{author}{Cacioppo, J.~T.}, \bibinfo{author}{Petty, R.~E.}, \bibinfo{author}{Losch, M.~E.} \& \bibinfo{author}{Kim, H.~S.}
\newblock \bibinfo{title}{Electromyographic activity over facial muscle regions can differentiate the valence and intensity of affective reactions.}
\newblock \emph{\bibinfo{journal}{Journal of Personality and Social Psychology}} \textbf{\bibinfo{volume}{50}}, \bibinfo{pages}{260--268} (\bibinfo{year}{1986}).

\bibitem{Mat2017}
\bibinfo{author}{Mattek, A.~M.}, \bibinfo{author}{Wolford, G.~L.} \& \bibinfo{author}{Whalen, P.~J.}
\newblock \bibinfo{title}{A mathematical model captures the structure of subjective affect}.
\newblock \emph{\bibinfo{journal}{Perspectives on Psychological Science}} \textbf{\bibinfo{volume}{12}}, \bibinfo{pages}{508--526} (\bibinfo{year}{2017}).
\newblock \urlprefix\url{https://doi.org/10.1177/1745691616685863}.

\bibitem{Pos2005}
\bibinfo{author}{Posner, J.}, \bibinfo{author}{Russell, J.~A.} \& \bibinfo{author}{Peterson, B.~S.}
\newblock \bibinfo{title}{The circumplex model of affect: An integrative approach to affective neuroscience, cognitive development, and psychopathology}.
\newblock \emph{\bibinfo{journal}{Development and Psychopathology}} \textbf{\bibinfo{volume}{17}} (\bibinfo{year}{2005}).
\newblock \urlprefix\url{https://doi.org/10.1017/s0954579405050340}.

\bibitem{Rus1980}
\bibinfo{author}{Russell, J.~A.}
\newblock \bibinfo{title}{A circumplex model of affect}.
\newblock \emph{\bibinfo{journal}{Journal of Personality and Social Psychology}} \textbf{\bibinfo{volume}{39}}, \bibinfo{pages}{1161--1178} (\bibinfo{year}{1980}).

\bibitem{Sch1952}
\bibinfo{author}{Schlosberg, H.}
\newblock \bibinfo{title}{The description of facial expressions in terms of two dimensions.}
\newblock \emph{\bibinfo{journal}{Journal of Experimental Psychology}} \textbf{\bibinfo{volume}{44}}, \bibinfo{pages}{229--237} (\bibinfo{year}{1952}).
\newblock \urlprefix\url{https://doi.org/10.1037/h0055778}.

\bibitem{Ash2005}
\bibinfo{editor}{Ashkanasy, N.~M.}, \bibinfo{editor}{Zerbe, W.~J.} \& \bibinfo{editor}{H\"{a}rtel, C.~E.} (eds) \emph{\bibinfo{title}{The Effect of Affect in Organizational Settings}}  (\bibinfo{publisher}{Emerald Group Publishing Limited}, \bibinfo{year}{2005}).
\newblock \urlprefix\url{https://doi.org/10.1016/s1746-9791(2005)1}.

\bibitem{Tru2015}
\bibinfo{author}{Trull, T.~J.}, \bibinfo{author}{Lane, S.~P.}, \bibinfo{author}{Koval, P.} \& \bibinfo{author}{Ebner-Priemer, U.~W.}
\newblock \bibinfo{title}{Affective dynamics in psychopathology}.
\newblock \emph{\bibinfo{journal}{Emotion Review}} \textbf{\bibinfo{volume}{7}}, \bibinfo{pages}{355--361} (\bibinfo{year}{2015}).
\newblock \urlprefix\url{https://doi.org/10.1177/1754073915590617}.

\bibitem{Wei1996}
\bibinfo{author}{Weiss, H.} \& \bibinfo{author}{Cropanzano, R.}
\newblock \bibinfo{title}{Affective events theory: a theoretical discussion of the structure, causes and consequences of affective experiences at work.}
\newblock \emph{\bibinfo{journal}{Research in Organizational Behavior}} \textbf{\bibinfo{volume}{8}}, \bibinfo{pages}{1--74} (\bibinfo{year}{1996}).

\bibitem{OpenBCI}
\bibinfo{title}{Openbci}.
\newblock \bibinfo{howpublished}{https://openbci.com/}.
\newblock \bibinfo{note}{Accessed: 2025-08-19}.

\bibitem{Jia2019}
\bibinfo{author}{L., J.}, \bibinfo{author}{A., S.} \& \bibinfo{author}{et~al., L.~D.}
\newblock \bibinfo{title}{Brainnet: A multi-person brain-to-brain interface for direct collaboration between brains.}
\newblock \emph{\bibinfo{journal}{Sci Rep}} \textbf{\bibinfo{volume}{117}}, \bibinfo{pages}{6115} (\bibinfo{year}{2019}).

\bibitem{Mon2023}
\bibinfo{author}{Montgomery, S.~M.}, \bibinfo{author}{Nair, N.}, \bibinfo{author}{Chen, P.} \& \bibinfo{author}{Dikker, S.}
\newblock \bibinfo{title}{Introducing emotibit, an open-source multi-modal sensor for measuring research-grade physiological signals}.
\newblock \emph{\bibinfo{journal}{Science Talks}} \textbf{\bibinfo{volume}{6}}, \bibinfo{pages}{100181} (\bibinfo{year}{2023}).

\bibitem{Emotibit}
\bibinfo{title}{Emotibit}.
\newblock \bibinfo{howpublished}{https://www.emotibit.com/}.
\newblock \bibinfo{note}{Accessed: 2025-08-19}.

\bibitem{Mic2018}
\bibinfo{author}{Michel, C.~M.} \& \bibinfo{author}{Koenig, T.}
\newblock \bibinfo{title}{Eeg microstates as a tool for studying the temporal dynamics of whole-brain neuronal networks: A review}.
\newblock \emph{\bibinfo{journal}{NeuroImage}} \textbf{\bibinfo{volume}{180}}, \bibinfo{pages}{577--593} (\bibinfo{year}{2018}).

\bibitem{Maa2024}
\bibinfo{author}{Maama, M.}, \bibinfo{author}{Ambrosio, B.}, \bibinfo{author}{Aziz-Alaoui, M.} \& \bibinfo{author}{Mintchev, S.~M.}
\newblock \bibinfo{title}{Emergent properties in a v1-inspired network of hodgkin–huxley neurons}.
\newblock \emph{\bibinfo{journal}{Mathematical Modelling of Natural Phenomena}} \textbf{\bibinfo{volume}{19}}, \bibinfo{pages}{3} (\bibinfo{year}{2024}).
\newblock \urlprefix\url{http://dx.doi.org/10.1051/mmnp/2024001}.

\bibitem{Amb2009}
\bibinfo{author}{Ambrosio, B.} \& \bibinfo{author}{Francoise, J.-P.}
\newblock \bibinfo{title}{Propagation of bursting oscillations}.
\newblock \emph{\bibinfo{journal}{Philosophical Transactions of the Royal Society A: Mathematical, Physical and Engineering Sciences}} \textbf{\bibinfo{volume}{367}}, \bibinfo{pages}{4863--4875} (\bibinfo{year}{2009}).

\bibitem{Amb2012}
\bibinfo{author}{Ambrosio, B.} \& \bibinfo{author}{Aziz-Alaoui, M.-A.}
\newblock \bibinfo{title}{Synchronization and control of coupled reaction-diffusion systems of the fitzhugh-nagumo type}.
\newblock \emph{\bibinfo{journal}{Computer and Mathematics with application}} \textbf{\bibinfo{volume}{64}}, \bibinfo{pages}{934--943} (\bibinfo{year}{2012}).

\bibitem{Amb2013}
\bibinfo{author}{Ambrosio, B.} \& \bibinfo{author}{Aziz-Alaoui, M.-A.}
\newblock \bibinfo{title}{Synchronization and control of a network of coupled reaction-diffusion systems of generalized fitzhugh-nagumo type}.
\newblock \emph{\bibinfo{journal}{ESAIM:Proceedings}} \textbf{\bibinfo{volume}{39}} (\bibinfo{year}{2013}).

\bibitem{Amb2016}
\bibinfo{author}{Ambrosio, B.} \& \bibinfo{author}{Aziz-Alaoui, M.~A.}
\newblock \bibinfo{title}{Basin of attraction of solutions with pattern formation in slow{\textendash}fast reaction{\textendash}diffusion systems}.
\newblock \emph{\bibinfo{journal}{Acta Biotheoretica}} \textbf{\bibinfo{volume}{64}}, \bibinfo{pages}{311--325} (\bibinfo{year}{2016}).
\newblock \urlprefix\url{https://doi.org/10.1007/s10441-016-9294-z}.

\bibitem{Amb2017}
\bibinfo{author}{Ambrosio, B.}
\newblock \bibinfo{title}{Hopf bifurcation in an oscillatory-excitable reaction{\textendash}diffusion model with spatial heterogeneity}.
\newblock \emph{\bibinfo{journal}{International Journal of Bifurcation and Chaos}} \textbf{\bibinfo{volume}{27}}, \bibinfo{pages}{1750065} (\bibinfo{year}{2017}).

\bibitem{Amb2017-2}
\bibinfo{author}{Ambrosio, B.}, \bibinfo{author}{Aziz-Alaoui, M.~A.} \& \bibinfo{author}{Balti, A.}
\newblock \bibinfo{title}{Propagation of bursting oscillations in coupled non-homogeneous hodgkin{\textendash}huxley reaction{\textendash}diffusion systems}.
\newblock \emph{\bibinfo{journal}{Differential Equations and Dynamical Systems}}  (\bibinfo{year}{2017}).
\newblock \urlprefix\url{https://doi.org/10.1007/s12591-017-0366-6}.

\bibitem{Amb2023}
\bibinfo{author}{Ambrosio, B.}
\newblock \bibinfo{title}{Qualitative analysis of certain reaction-diffusion systems of the fitzhugh-nagumo type}.
\newblock \emph{\bibinfo{journal}{Evolution Equations and Control Theory}} \textbf{\bibinfo{volume}{12}}, \bibinfo{pages}{1507--1526} (\bibinfo{year}{2023}).

\bibitem{Amb2024}
\bibinfo{author}{Ambrosio, B.}, \bibinfo{author}{Aziz-Alaoui, M.~A.} \& \bibinfo{author}{Oujbara, A.}
\newblock \bibinfo{title}{Synchronization in a three level network of all-to-all periodically forced hodgkin–huxley reaction–diffusion equations}.
\newblock \emph{\bibinfo{journal}{Mathematics}} \textbf{\bibinfo{volume}{12}}, \bibinfo{pages}{1382} (\bibinfo{year}{2024}).
\newblock \urlprefix\url{http://dx.doi.org/10.3390/math12091382}.

\bibitem{Xi2022}
\bibinfo{author}{Xi, H.} \& \bibinfo{author}{Su, J.}
\newblock \bibinfo{title}{A harmonic function method for eeg source reconstruction}.
\newblock \emph{\bibinfo{journal}{Electronic Research Archiv}} \textbf{\bibinfo{volume}{30}}, \bibinfo{pages}{492--514} (\bibinfo{year}{2022}).

\bibitem{Amb2022}
\bibinfo{author}{Ambrosio, B.} \& \bibinfo{author}{Young, L.-S.}
\newblock \bibinfo{title}{The use of reduced models to generate irregular, broad-band signals that resemble brain rhythms}.
\newblock \emph{\bibinfo{journal}{Frontiers in Computational Neuroscience}} \textbf{\bibinfo{volume}{16}} (\bibinfo{year}{2022}).
\newblock \urlprefix\url{http://dx.doi.org/10.3389/fncom.2022.889235}.

\bibitem{Vol2024}
\bibinfo{author}{Volpert, V.}, \bibinfo{author}{Sadaka, G.}, \bibinfo{author}{Mesnildrey, Q.} \& \bibinfo{author}{Beuter, A.}
\newblock \bibinfo{title}{Modelling eeg dynamics with brain sources}.
\newblock \emph{\bibinfo{journal}{Symmetry}} \textbf{\bibinfo{volume}{16}}, \bibinfo{pages}{189} (\bibinfo{year}{2024}).

\bibitem{Fra2025}
\bibinfo{author}{Frank, L.~R.} \emph{et~al.}
\newblock \bibinfo{title}{Imaging of brain electric field networks with spatially resolved eeg} \textbf{\bibinfo{volume}{13}}, \bibinfo{pages}{RP100123} (\bibinfo{year}{2025}).

\bibitem{Fit1961}
\bibinfo{author}{FitzHugh, R.}
\newblock \bibinfo{title}{Impulses and physiological states in theoretical models of nerve membrane}.
\newblock \emph{\bibinfo{journal}{Biophysical Journal}} \textbf{\bibinfo{volume}{1}}, \bibinfo{pages}{445--466} (\bibinfo{year}{1961}).

\bibitem{Nag1962}
\bibinfo{author}{Nagumo, J.}, \bibinfo{author}{Arimoto, S.} \& \bibinfo{author}{Yoshizawa, S.}
\newblock \bibinfo{title}{An active pulse transmission line simulating nerve axon}.
\newblock \emph{\bibinfo{journal}{Proceedings of the IRE}} \textbf{\bibinfo{volume}{50}}, \bibinfo{pages}{2061--2070} (\bibinfo{year}{1962}).

\bibitem{Fen1979}
\bibinfo{author}{Fenichel, N.}
\newblock \bibinfo{title}{Geometric singular perturbation theory for ordinary differential equations}.
\newblock \emph{\bibinfo{journal}{Journal of Differential Equations}} \textbf{\bibinfo{volume}{31}}, \bibinfo{pages}{53--98} (\bibinfo{year}{1979}).

\bibitem{Jon1995}
\bibinfo{author}{Jones, C. K. R.~T.}
\newblock \bibinfo{title}{Geometric singular perturbation theory}.
\newblock \emph{\bibinfo{journal}{Dynamical Systems (Montecatini Terme, 1994), Lecture Notes in Math.}} \textbf{\bibinfo{volume}{1609}}, \bibinfo{pages}{44--118} (\bibinfo{year}{1995}).

\bibitem{Kue2015}
\bibinfo{author}{Kuehn, C.}
\newblock \emph{\bibinfo{title}{Multiple Time Scale Dynamics}}  (\bibinfo{publisher}{Springer}, \bibinfo{year}{2015}).

\bibitem{Rin1985}
\bibinfo{author}{Rinzel, J.}
\newblock \bibinfo{title}{Excitation dynamics: Insights from simplified models}.
\newblock \emph{\bibinfo{journal}{Mathematical Topics in Population Biology, Morphogenesis and Neurosciences}} \textbf{\bibinfo{volume}{71}}, \bibinfo{pages}{267--314} (\bibinfo{year}{1985}).

\bibitem{Amb2021-2}
\bibinfo{author}{Ambrosio, B.} \& \bibinfo{author}{Mintchev, S.~M.}
\newblock \bibinfo{title}{Periodically kicked feedforward chains of simple excitable fitzhugh-nagumo neurons}.
\newblock \emph{\bibinfo{journal}{Nonlinear Dynamics}} \textbf{\bibinfo{volume}{110}}, \bibinfo{pages}{2805--2829} (\bibinfo{year}{2022}).

\bibitem{fre2012}
\bibinfo{author}{Freidlin, M.~I.} \& \bibinfo{author}{Wentzell, A.~D.}
\newblock \emph{\bibinfo{title}{Random Perturbations of Dynamical Systems}} \bibinfo{edition}{3rd} edn, Vol. \bibinfo{volume}{260} of \emph{\bibinfo{series}{Grundlehren der mathematischen Wissenschaften}} (\bibinfo{publisher}{Springer}, \bibinfo{year}{2012}).

\bibitem{ber2004}
\bibinfo{author}{Berglund, N.} \& \bibinfo{author}{Gentz, B.}
\newblock \bibinfo{title}{Noise-induced phenomena in slow–fast dynamical systems}.
\newblock \emph{\bibinfo{journal}{Stochastic Processes and their Applications}} \textbf{\bibinfo{volume}{113}}, \bibinfo{pages}{1--40} (\bibinfo{year}{2004}).

\bibitem{ber2006}
\bibinfo{author}{Berglund, N.} \& \bibinfo{author}{Gentz, B.}
\newblock \emph{\bibinfo{title}{Noise-Induced Phenomena in Slow–Fast Dynamical Systems: A Sample-Paths Approach}} Probability and Its Applications (\bibinfo{publisher}{Springer}, \bibinfo{year}{2006}).

\bibitem{wei2004}
\bibinfo{author}{E, W.}, \bibinfo{author}{Ren, W.} \& \bibinfo{author}{Vanden-Eijnden, E.}
\newblock \bibinfo{title}{Minimum action method for the study of rare events}.
\newblock \emph{\bibinfo{journal}{Communications on Pure and Applied Mathematics}} \textbf{\bibinfo{volume}{57}}, \bibinfo{pages}{637--656} (\bibinfo{year}{2004}).

\bibitem{hey2008}
\bibinfo{author}{Heymann, M.} \& \bibinfo{author}{Vanden-Eijnden, E.}
\newblock \bibinfo{title}{The geometric minimum action method: A least action principle on the space of curves}.
\newblock \emph{\bibinfo{journal}{Communications on Pure and Applied Mathematics}} \textbf{\bibinfo{volume}{61}}, \bibinfo{pages}{1052--1117} (\bibinfo{year}{2008}).

\bibitem{hey2015}
\bibinfo{author}{Heymann, M.} \& \bibinfo{author}{Vanden-Eijnden, E.}
\newblock \bibinfo{title}{Computing quasipotentials for nongradient sdes in 2d}.
\newblock \emph{\bibinfo{journal}{Communications in Mathematical Sciences}} \textbf{\bibinfo{volume}{13}}, \bibinfo{pages}{903--924} (\bibinfo{year}{2015}).

\bibitem{kra1940}
\bibinfo{author}{Kramers, H.~A.}
\newblock \bibinfo{title}{Brownian motion in a field of force and the diffusion model of chemical reactions}.
\newblock \emph{\bibinfo{journal}{Physica}} \textbf{\bibinfo{volume}{7}}, \bibinfo{pages}{284--304} (\bibinfo{year}{1940}).

\bibitem{han1990}
\bibinfo{author}{H{\"a}nggi, P.}, \bibinfo{author}{Talkner, P.} \& \bibinfo{author}{Borkovec, M.}
\newblock \emph{\bibinfo{title}{Reaction-rate theory: fifty years after Kramers}} Vol.~\bibinfo{volume}{62} (\bibinfo{year}{1990}).

\bibitem{nei1987}
\bibinfo{author}{Neishtadt, A.~I.}
\newblock \bibinfo{title}{Persistence of stability loss for dynamical bifurcations. i}.
\newblock \emph{\bibinfo{journal}{Differential Equations}} \textbf{\bibinfo{volume}{23}}, \bibinfo{pages}{1385--1391} (\bibinfo{year}{1987}).

\bibitem{rit2016}
\bibinfo{author}{Ritchie, P.} \& \bibinfo{author}{Sieber, J.}
\newblock \bibinfo{title}{The effective potential for quasi-stationary distributions of fast–slow stochastic systems}.
\newblock \emph{\bibinfo{journal}{SIAM Journal on Applied Dynamical Systems}} \textbf{\bibinfo{volume}{15}}, \bibinfo{pages}{601--623} (\bibinfo{year}{2016}).

\bibitem{leh1987}
\bibinfo{author}{Lehmann, D.}
\newblock \bibinfo{title}{Eeg microstates: theory and practice}.
\newblock \emph{\bibinfo{journal}{Electroencephalography and Clinical Neurophysiology}} \textbf{\bibinfo{volume}{67}}, \bibinfo{pages}{271--288} (\bibinfo{year}{1987}).

\bibitem{hay2025}
\bibinfo{author}{Haydock, D.}
\newblock \bibinfo{title}{Eeg microstate syntax analysis: A review}.
\newblock \emph{\bibinfo{journal}{NeuroImage}}  (\bibinfo{year}{2025}).
\newblock \bibinfo{note}{Https://pubmed.ncbi.nlm.nih.gov/39961498/}.

\bibitem{kas2025}
\bibinfo{author}{Kashihara, S.} \& \bibinfo{author}{colleagues}.
\newblock \bibinfo{title}{Topographical polarity reveals continuous eeg microstate transitions}.
\newblock \emph{\bibinfo{journal}{Cerebral Cortex}}  (\bibinfo{year}{2025}).
\newblock \bibinfo{note}{Https://pubmed.ncbi.nlm.nih.gov/40794889/}.

\bibitem{Das2023}
\bibinfo{author}{Das, S.} \emph{et~al.}
\newblock \emph{\bibinfo{title}{Movie’s-Emotracker: Movie Induced Emotion Detection by Using EEG and AI Tools}}, \bibinfo{pages}{583–595} (\bibinfo{publisher}{Springer Nature Singapore}, \bibinfo{year}{2023}).

\bibitem{Liu2023}
\bibinfo{author}{Liu, J.} \emph{et~al.}
\newblock \bibinfo{title}{The eeg microstate representation of discrete emotions}.
\newblock \emph{\bibinfo{journal}{International Journal of Psychophysiology}} \textbf{\bibinfo{volume}{186}}, \bibinfo{pages}{33--41} (\bibinfo{year}{2023}).

\end{thebibliography}

 %for arXiv submission

\end{document}